# Constructions for $q$-Ary Constant-Weight Codes

Yeow Meng Chee and San Ling

*Abstract*—This paper introduces a new combinatorial construction for $q$-ary constant-weight codes which yields several families of optimal codes and asymptotically optimal codes. The construction reveals intimate connection between $q$-ary constant-weight codes and sets of pairwise disjoint combinatorial designs of various types.

## I. INTRODUCTION

THE class of $q$-ary constant-weight codes (all terms are defined in the next section) has attracted some recent attention due to several important applications requiring nonbinary alphabets, such as coding for bandwidth-efficient channels and design of oligonucleotide sequences for DNA computing. While a vast amount of knowledge exists for binary constant-weight codes [1], relatively little is known about $q$-ary constant-weight codes when $q > 2$. As with binary codes, the interest is in determining $A_q(n, d, w)$, the maximum size of an $(n, d, w)_q$-code. We briefly summarize some past work as follows.

  i) General constructions for $(n, d, w)_q$-codes are studied in [2]–[4].
 ii) $A_q(n, 3, 3)$ is studied in [2], [5]–[14].
iii) $A_q(n, 5, 4)$ is studied in [15], [16].
 iv) $A_3(n, d, w)$ is studied in [17]–[24].
  v) $A_4(n, d, w)$ is studied in [25].

Most of these known constructions apply to very constrained parameters, focusing on fixed $q$, $n \in \{q-1, q, q+1\}$, $n$ a prime power, or $(d, w) = (3, 3)$. The number-theoretic constraints on $n$ and $q$ arise because of the algebraic constructions considered. Our approach in this paper is combinatorial.

We introduce a new general construction for $q$-ary constant-weight codes from binary constant-weight codes that yields several families of optimal and asymptotically optimal $(n, d, w)_q$-codes. In particular, we completely determine:

  i) the exact value of $A_q(n, 3, 2)$ for all $q$ and $n$;
 ii) the exact value of $A_3(n, 4, 3)$ for all $n$;
iii) the exact value of $A_q(n, 4, 3)$ for all $q$ and $n \equiv 0, 1, 2,$ or $3 \pmod 6$;
 iv) the asymptotic value of $A_q(n, 4, 3)$ for all $q$; and
  v) asymptotic lower bounds for $A_q(n, d, w)$, within a factor of $q^\epsilon$ from optimal, for any $\epsilon > 0$.

Our construction shows intimate connections between $(n, d, w)_q$-codes and sets of pairwise disjoint combinatorial designs of various types, including packings, $t$-designs, and group divisible designs.

We also give a new probabilistic construction for $(n, w+1, w)_q$-codes that is better than the $q$-ary Gilbert–Varshamov bound when $w$ is even and $q \leq 4$.

The outline of the paper is as follows. In Section II, basic notions and results in coding theory and combinatorial tools used in the paper are discussed. Section III outlines the main strategy used in the paper for constructing optimal nonbinary constant-weight codes. In Section IV, the value of $A_q(n, 3, 2)$ is determined completely. As some further results in combinatorial design theory are needed for the determination of $A_3(n, 4, 3)$ for some values of $n$, these new results are contained in Section V. Section VI is devoted to the determination of $A_q(n, 4, 3)$. Exact values of $A_q(n, 4, 3)$ for all $q$ and $n \equiv 0, 1, 2,$ or $3 \pmod 6$, as well as the values of $A_3(n, 4, 3)$ for all $n$ and the asymptotic value of $A_q(n, 4, 3)$ for all $q$, are obtained. In Section VII, the main strategy is applied to determine $A_q(13, 6, 4)$. In Section VIII, we consider the problem of determining the values of $A_q(n, w+1, w)$, and obtain some bounds for these values. Then, in Section IX, a probabilistic construction for $(n, d, w)_q$-codes is given, and bounds for the values of $A_q(n, d, w)$ are obtained. Finally, a brief conclusion is given in Section X.

## II. DEFINITIONS AND NOTATIONS

In this section, we recall some basic notions related to constant-weight codes. As combinatorial objects, such as set systems, designs, packings, and graphs, play an instrumental role in many of the proofs in this paper, we also recall some of the relevant definitions and results for these objects.

### A. $q$-Ary Constant-Weight Codes

The set of integers $\{1, \ldots, n\}$ is denoted by $[n]$. For $q$ a positive integer, we denote the ring $\mathbb{Z}/q\mathbb{Z}$ by $\mathbb{Z}_q$. The $i$th coordinate of a vector $\mathsf{u}$ is denoted by $\mathsf{u}_i$. The $q$-ary Hamming $n$-space is the set $\mathcal{H}_q(n) = \mathbb{Z}_q^n$ endowed with the Hamming distance metric $d_H$ defined as follows:

$$d_H(\mathsf{u}, \mathsf{v}) = |\{1 \leq i \leq n : \mathsf{u}_i \neq \mathsf{v}_i\}|,$$

the number of coordinates where $\mathsf{u}$ and $\mathsf{v}$ differ. The *Hamming weight* of a vector $\mathsf{u} \in \mathcal{H}_q(n)$ is the quantity $d_H(\mathsf{u}, 0)$, the number of nonzero coordinates of $\mathsf{u}$. The *support* of $\mathsf{u}$ is defined to be the set $\{i : \mathsf{u}_i \neq 0\}$. In other words, the Hamming

Manuscript received August 2, 2006; revised October 11, 2006. The research of S. Ling was supported in part by NTU under Research Grant M58110003.
Y. M. Chee is with the Interactive and Digital Medias R&D Program Office, Media Development Authority, Singapore 179369. He is also with the Division of Mathematical Sciences, School of Physical and Mathematical Sciences, Nanyang Technological University, Singapore 637616 (e-mail: ymchee@alumni.uwaterloo.ca).
S. Ling is with the Division of Mathematical Sciences, School of Physical and Mathematical Sciences, Nanyang Technological University, Singapore 637616 (e-mail: lingsan@ntu.edu.sg).
Communicated by I. Dumer, Associate Editor for Coding Theory.
Digital Object Identifier 10.1109/TIT.2006.887499

weight of u is the size of the support of u. The set of all elements in $\mathcal{H}_q(n)$ having Hamming weight $w$ is denoted $\mathcal{H}_q(n,w)$. A $q$-ary code of length $n$, distance $d$, and constant weight $w$, denoted $(n,d,w)_q$-*code*, is a nonempty set $\mathcal{C} \subseteq \mathcal{H}_q(n,w)$ such that $d_H(\mathsf{u},\mathsf{v}) \geq d$ for all $\mathsf{u},\mathsf{v} \in \mathcal{C}$, $\mathsf{u} \neq \mathsf{v}$. The elements of $\mathcal{C}$ are called *codewords*.

The number of codewords in an $(n,d,w)_q$-code is called the *size* of the code. The maximum size of an $(n,d,w)_q$-code is denoted $A_q(n,d,w)$. An $(n,d,w)_q$-code having $A_q(n,d,w)$ codewords is said to be *optimal*.

Let $\{\mathcal{C}_n : n \geq 1\}$ be a family of codes such that $\mathcal{C}_n$ is an $(n,d,w)_q$-code. Then $\{\mathcal{C}_n : n \geq 1\}$ is said to be *asymptotically optimal* if $|\mathcal{C}_n| \sim A_q(n,d,w)$, where $f(n) \sim g(n)$ means $\lim_{n \to \infty} f(n)/g(n) = 1$. We also write $f(n) \gtrsim g(n)$ if $\lim_{n \to \infty} f(n)/g(n) \geq 1$.

The following bounds have been established by Svanström [19].

*Lemma 1 (Svanström [19]):*
$$A_q(n,d,w) \leq \left\lfloor \frac{n}{n-w} A_q(n-1,d,w) \right\rfloor.$$

*Lemma 2 (Svanström [19]):*
$$A_q(n,d,w) \leq \left\lfloor \frac{n(q-1)}{w} A_q(n-1,d,w-1) \right\rfloor.$$

*B. Set Systems, Designs, and Packings*

A *set system* is a pair $(X,\mathcal{A})$ such that $X$ is a finite set of *points* and $\mathcal{A}$ is a set of subsets of $X$, called *blocks*. The number of points, $|X|$, is the *order* of the set system. Let $K$ be a set of positive integers. A set system $(X,\mathcal{A})$ is said to be $K$-*uniform* if $|A| \in K$ for all $A \in \mathcal{A}$. Two set systems $(X,\mathcal{A})$ and $(X,\mathcal{B})$ are *disjoint* if $\mathcal{A} \cap \mathcal{B} = \varnothing$.

Given a set system $([n],\mathcal{A})$, and $A \in \mathcal{A}$, let $\iota(A) \in \{0,1\}^n$ denote the incidence vector such that
$$\iota(A)_i = \begin{cases} 1, & \text{if } i \in A \\ 0, & \text{otherwise.} \end{cases}$$

The set $\{\iota(A) : A \in \mathcal{A}\} \subseteq \mathcal{H}_2(n)$ is called *the code of the set system* $([n],\mathcal{A})$.

Let $(X,\mathcal{A})$ be a set system and $\mathcal{G} = \{G_1,\ldots,G_s\}$ be a partition of $X$ into subsets, called *groups*. The triple $(X,\mathcal{G},\mathcal{A})$ is a *group divisible design* (GDD) when every 2-subset of $X$ not contained in a group appears in exactly one block and $|A \cap G| \leq 1$ for all $A \in \mathcal{A}$ and $G \in \mathcal{G}$. We denote a GDD $(X,\mathcal{G},\mathcal{A})$ by $K$-GDD if $(X,\mathcal{A})$ is $K$-uniform. The *type* of a GDD $(X,\mathcal{G},\mathcal{A})$ is the multiset $[|G| : G \in \mathcal{G}]$. We use the exponential notation to describe the type of a GDD: a GDD of type $g_1^{t_1} \cdots g_s^{t_s}$ is a GDD where there are exactly $t_i$ groups of size $g_i$, $1 \leq i \leq s$. When a GDD $(X,\mathcal{G},\mathcal{A})$ has all groups of size one (that is, of type $1^t$), it is common to identify the GDD simply with the set system $(X,\mathcal{A})$.

A $\{3\}$-GDD of type $1^n$ is known as a *Steiner triple system of order* $n$, and is denoted $\mathrm{STS}(n)$.

*Theorem 1 (Folklore, See [26]):* There exists an $\mathrm{STS}(n)$ if and only if $n \equiv 1$ or $3 \pmod 6$.

*Theorem 2 (Fort and Hedlund [27]):* There exists a $\{3\}$-GDD of type $5^1 1^{6t}$ for all $t \geq 0$.

Two GDDs, $(X,\mathcal{G},\mathcal{A})$ and $(X,\mathcal{G},\mathcal{B})$, are said to have *intersection size* $m$ if $|\mathcal{A} \cap \mathcal{B}| = m$. Two GDDs with intersection size zero are *disjoint*. The set of all intersection sizes of two $\{3\}$-GDDs of type $g^t$ is denoted $\mathrm{Int}(g^t)$. The following results on the intersection sizes of $\{3\}$-GDDs are useful.

*Theorem 3 (Butler and Hoffman [28]):* Let $t \geq 3$, $g^2 \binom{t}{2} \equiv 0 \pmod 3$, and $g(t-1) \equiv 0 \pmod 2$. Let $b(g^t) = g^2 \binom{t}{2}/3$ (the number of blocks in a $\{3\}$-GDD of type $g^t$) and define

$$\mathcal{I}(g^t) = \{0,1,\ldots,b(g^t)\} \setminus \{b(g^t)-5, b(g^t)-3, b(g^t)-2, b(g^t)-1\}.$$

Then $\mathrm{Int}(g^t) = \mathcal{I}(g^t)$, except that

i) $\mathrm{Int}(1^9) = \mathcal{I}(1^9) \setminus \{5,8\}$;
ii) $\mathrm{Int}(2^4) = \mathcal{I}(2^4) \setminus \{1,4\}$;
iii) $\mathrm{Int}(3^3) = \mathcal{I}(3^3) \setminus \{1,2,5\}$;
iv) $\mathrm{Int}(4^3) = \mathcal{I}(4^3) \setminus \{5,7,10\}$.

*Theorem 4 (Chee [29]):* Let $3t + r \equiv 1$ or $3 \pmod 6$. Then there exists a pair of disjoint $\{3\}$-GDDs of type $3^t 1^r$ if and only if one of the following conditions is satisfied:

i) $t = 0$;
ii) $3t + r \geq 13$; or
iii) $(t,r) \in \{(1,4),(1,6),(3,0)\}$.

An *incomplete $K$-GDD* ($K$-IGDD) of type $(g,h)^t$ is a quadruple $(X,\mathcal{G},\mathcal{F},\mathcal{A})$ such that $\mathcal{G}$ is a partition of $X$ into $t$ *groups*, each of size $g$, $\mathcal{F} = \{F_1,\ldots,F_t\}$ is a set of $h$-subsets of $X$ (called *holes*) such that $F_i \subseteq G_i$, and $(X,\mathcal{A})$ is a $K$-uniform set system which satisfies the properties:

i) any 2-subset of $X$ contained in a group is not contained in any blocks of $\mathcal{A}$;
ii) if a 2-subset of $X$ is contained in $\cup_{i=1}^t F_i$, then it is not contained in any blocks of $\mathcal{A}$; and
iii) any other 2-subset of $X$ is contained in exactly one block of $\mathcal{A}$.

A *Latin square of side* $n$ is an $n \times n$ array in which each cell contains an element from $[n]$, such that each element of $[n]$ occurs exactly once in each row and exactly once in each column. If in a Latin square $L$ of side $n$, the $k^2$ cells defined by $k$ rows and $k$ columns, form a Latin square of side $k$, it is a *subsquare (of $L$) of side $k$*. Two Latin squares $L$ and $L'$ have a *common subsquare* of side $k$ if the $k$ rows and $k$ columns defining a subsquare $S$ of side $k$ in $L$ also define a subsquare $S'$ of side $k$ in $L'$, and furthermore, $S = S'$.

*Lemma 3 (Evans [30]):* A Latin square of side $n$ with a subsquare of side $k$ exists if and only if $k \leq \lfloor n/2 \rfloor$.

Two Latin squares $L$ and $L'$ of side $n$ having a common subsquare $S$ are said to be *disjoint* if for all $i,j \in [n]$, the $(i,j)$th entries of $L$ and $L'$ are different, except when the entries are in $S$.

A $t$-$(n,k,\lambda)$-*packing* (resp., *design*) is a $\{k\}$-uniform set system $(X,\mathcal{A})$ of order $n$ such that every $t$-subset of $X$ is contained in at most (resp., exactly) $\lambda$ blocks of $\mathcal{A}$. When $\lambda = 1$, such a packing (resp., design) is also sometimes called a $(t, k, n)$-packing (resp., design). A $2$-$(n, 3, 1)$ design is none other than a Steiner triple system of order $n$ defined above. The maximum number of blocks in a $t$-$(n, k, \lambda)$-packing is denoted $D_\lambda(n, k, t)$. Any $t$-$(n, k, \lambda)$-packing with $D_\lambda(n, k, t)$ blocks is called *maximum*. The following upper bound on $D_\lambda(n, k, t)$ is due to Johnson [31] and Schönheim [32]:

*Lemma 4 (Johnson, Schönheim):*

$$D_\lambda(n, k, t) \leq \left\lfloor \frac{n}{k} \left\lfloor \frac{n-1}{k-1} \cdots \left\lfloor \frac{n-t+1}{k-t+1} \lambda \right\rfloor \cdots \right\rfloor \right\rfloor \leq \frac{\lambda \binom{n}{t}}{\binom{k}{t}}.$$

There is an intimate relationship between packings and binary constant-weight codes.

*Lemma 5:* The code of a $(t, k, n)$-packing is an $(n, 2(k-t+1), k)_2$-code, and *vice versa*.

Next, we recall the notion of large sets.

A *large set* of $t$-$(n, k, \lambda)$ designs is a set

$$\{(X, \mathcal{A}_1), (X, \mathcal{A}_2), \ldots, (X, \mathcal{A}_m)\}$$

of $m = \binom{n-t}{k-t}/\lambda$ pairwise disjoint $t$-$(n, k, \lambda)$ designs such that $\cup_{i=1}^m \mathcal{A}_i = \binom{X}{k}$, the set of all $k$-subsets of $X$.

Let $n \equiv 5 \pmod{6}$. A *large set* of maximum $(2, 3, n)$-packings is a set $\{(X, \mathcal{A}_1), (X, \mathcal{A}_2), \ldots, (X, \mathcal{A}_{n-4})\}$ of $n-4$ pairwise disjoint maximum $(2, 3, n)$-packings.

The following results on the existence of large sets will be used later to construct optimal $q$-ary constant-weight codes.

*Theorem 5: (Lu [33]–[38], Teirlinck [39]):* There exists a large set of STS$(n)$ if and only if $n \equiv 1$ or $3 \pmod{6}$, $n \neq 7$. There exist two disjoint STS$(7)$.

A new and simpler proof for Theorem 5 was recently obtained by Ji [40].

*Theorem 6: (Cao, Ji, and Zhu [41]):* There exists a large set of maximum $(2, 3, n)$-packings for $n \equiv 5 \pmod{6}$.

*Theorem 7 (Chouinard [42]):* A large set of $2$-$(13, 4, 1)$ designs exists.

### C. Graphs and Factorizations

A *graph* $G = (V, E)$ consists of a set $V$ of *vertices* together with a set $E$ of *edges*, where an edge may be considered as a set consisting of exactly two vertices in $V$ (hence, a graph is a $\{2\}$-uniform set system). For any positive integer $n$, the *complete graph* $K_n$ is the graph $(V, \binom{V}{2})$.

For any graph $G = (V, E)$, a *one-factor* is a subset of $E$ in which every vertex in $V$ appears in precisely one edge. A *one-factorization* of $G$ is a set of one-factors that partitions $E$. A *near-one-factor* is a subset of $E$ in which every vertex of $V$, except for one, appears in precisely one edge, while the remaining vertex is isolated. A partition of the edge-set $E$ into near-one-factors is called a *near-one-factorization*. In any near-one-factorization of $K_n$, every vertex appears in exactly one near-one-factor as an isolated vertex.

The following is known.

*Theorem 8 (Folklore, See [43]):* There exists a one-factorization of $K_n$ whenever $n$ is even, and a near-one-factorization of $K_n$ whenever $n$ is odd.

## III. A GENERAL STRATEGY

The following strategy and its variations are used several times in this paper to construct $(n, d, w)_q$-codes.

For $1 \leq s \leq q-1$, suppose there exist $s$ distinct binary constant-weight codes, say, $\mathcal{C}_1, \mathcal{C}_2, \ldots, \mathcal{C}_s$, of length $n$, distance $d'$, and weight $w$, such that $\mathcal{C}_i \cap \mathcal{C}_j = \varnothing$, $1 \leq i < j \leq s$. On $\mathcal{C}_i$, replace each occurrence of $1$ in each codeword by $i$ to yield the $q$-ary code $\mathcal{C}'_i$. Then $\mathcal{C}' = \cup_{i=1}^s \mathcal{C}'_i$ is a $q$-ary code of constant weight $w$. It is also obvious that $\mathcal{C}'$ has $\sum_{i=1}^s |\mathcal{C}_i|$ codewords. In particular, if $\mathcal{C}_1, \ldots, \mathcal{C}_s$ are optimal binary constant-weight codes, then it follows that $|\mathcal{C}'| = s A_2(n, d', w)$. The distance $d$ of $\mathcal{C}'$ can also be determined, often through a simple combinatorial argument. This construction therefore gives the lower bound

$$A_q(n, d, w) \geq s A_2(n, d', w).$$

In most of the proofs in the remainder of this paper, the binary constant-weight codes $\mathcal{C}_1, \ldots, \mathcal{C}_s$ arise either from a large set of certain $2$-$(n, k, \lambda)$ designs or packings, or from some partition of a binary Johnson space.

## IV. $A_q(n, 3, 2)$

In this section, we determine the exact values of $A_q(n, 3, 2)$ for all $q$ and $n$.

*Theorem 9:* For all $q > 1$ and positive integers $n$, we have

$$A_q(n, 3, 2) = \begin{cases} \left\lfloor \frac{(q-1)n}{2} \right\rfloor, & \text{if } q \leq n \\ \binom{n}{2}, & \text{if } q > n. \end{cases}$$

*Proof:* When $n$ is even, let $\{F_1, \ldots, F_{n-1}\}$ be a one-factorization of the complete graph $K_n$. For each $1 \leq i \leq n-1$, the code $\mathcal{C}_i$ corresponding to the one-factor $F_i$ is easily seen to be an $(n, 4, 2)_2$-code of size $n/2$. Clearly, $\mathcal{C}_i \cap \mathcal{C}_j = \varnothing$ for all $1 \leq i < j \leq n-1$.

When $n$ is odd, let $\{F_1, \ldots, F_n\}$ be a near-one-factorization of the complete graph $K_n$, and let $\mathcal{C}_i$ denote the code of $F_i$. Without loss of generality, we may label the near-one-factors and the vertices in such a way that the vertex $v_i$ is the unique vertex not in the support of $\mathcal{C}_i$ and that the code $\mathcal{C}_n$ consists of the $(n-1)/2$ codewords

$$(1, 1, 0, \ldots, 0), (0, 0, 1, 1, 0, \ldots, 0), \ldots, (0, \ldots, 0, 1, 1, 0).$$

Hence, for $1 \leq i \leq n$, each $\mathcal{C}_i$ is an $(n, 4, 2)_2$-code of size $(n-1)/2$ and $\mathcal{C}_i \cap \mathcal{C}_j = \varnothing$ for $1 \leq i < j \leq n$.

Assume that $q \leq n$.

TABLE I
A PAIR OF DISJOINT $\{3\}$-GDDS OF TYPE $5^1 1^6$

| blocks | blocks |
|---|---|
| $\{0,5,10\}$ $\{0,6,9\}$ $\{0,7,8\}$ | $\{0,5,9\}$ $\{0,6,8\}$ $\{0,7,10\}$ |
| $\{1,5,9\}$ $\{1,6,7\}$ $\{1,8,10\}$ | $\{1,5,6\}$ $\{1,7,8\}$ $\{1,9,10\}$ |
| $\{2,5,7\}$ $\{2,6,8\}$ $\{2,9,10\}$ | $\{2,5,8\}$ $\{2,6,10\}$ $\{2,7,9\}$ |
| $\{3,5,6\}$ $\{3,7,10\}$ $\{3,8,9\}$ | $\{3,5,7\}$ $\{3,6,9\}$ $\{3,8,10\}$ |
| $\{4,5,8\}$ $\{4,6,10\}$ $\{4,7,9\}$ | $\{4,5,10\}$ $\{4,6,7\}$ $\{4,8,9\}$ |

TABLE II
A PAIR OF DISJOINT $\{3\}$-GDDS OF TYPE $5^1 1^{12}$

| blocks | | blocks | |
|---|---|---|---|
| $\{0,5,11\}$ | $\{0,6,10\}$ | $\{0,5,13\}$ | $\{0,6,15\}$ |
| $\{0,7,13\}$ | $\{0,8,16\}$ | $\{0,7,16\}$ | $\{0,8,10\}$ |
| $\{0,9,14\}$ | $\{0,12,15\}$ | $\{0,9,11\}$ | $\{0,12,14\}$ |
| $\{1,10,12\}$ | $\{1,11,16\}$ | $\{1,5,9\}$ | $\{1,6,12\}$ |
| $\{1,13,15\}$ | $\{1,5,8\}$ | $\{1,7,15\}$ | $\{1,8,11\}$ |
| $\{1,6,14\}$ | $\{1,7,9\}$ | $\{1,10,14\}$ | $\{1,13,16\}$ |
| $\{2,5,15\}$ | $\{2,6,13\}$ | $\{2,5,14\}$ | $\{2,6,16\}$ |
| $\{2,7,11\}$ | $\{2,8,10\}$ | $\{2,7,12\}$ | $\{2,8,9\}$ |
| $\{2,9,12\}$ | $\{2,14,16\}$ | $\{2,10,11\}$ | $\{2,13,15\}$ |
| $\{3,5,12\}$ | $\{3,6,7\}$ | $\{3,5,11\}$ | $\{3,6,14\}$ |
| $\{3,8,14\}$ | $\{3,9,16\}$ | $\{3,7,9\}$ | $\{3,8,15\}$ |
| $\{3,10,13\}$ | $\{3,11,15\}$ | $\{3,10,16\}$ | $\{3,12,13\}$ |
| $\{4,5,13\}$ | $\{4,6,16\}$ | $\{4,5,7\}$ | $\{4,6,9\}$ |
| $\{4,7,10\}$ | $\{4,8,11\}$ | $\{4,8,13\}$ | $\{4,10,12\}$ |
| $\{4,9,15\}$ | $\{4,12,14\}$ | $\{4,11,16\}$ | $\{4,14,15\}$ |
| $\{5,6,9\}$ | $\{5,7,16\}$ | $\{5,6,10\}$ | $\{5,8,12\}$ |
| $\{5,10,14\}$ | $\{6,8,15\}$ | $\{5,15,16\}$ | $\{6,7,8\}$ |
| $\{6,11,12\}$ | $\{7,8,12\}$ | $\{6,11,13\}$ | $\{7,10,13\}$ |
| $\{7,14,15\}$ | $\{8,9,13\}$ | $\{7,11,14\}$ | $\{8,14,16\}$ |
| $\{9,10,11\}$ | $\{10,15,16\}$ | $\{9,10,15\}$ | $\{9,12,16\}$ |
| $\{11,13,14\}$ | $\{12,13,16\}$ | $\{9,13,14\}$ | $\{11,12,15\}$ |

From Lemma 2, we have

$$A_q(n,3,2) \leq \left\lfloor \frac{(q-1)n}{2} \right\rfloor.$$

Hence, it suffices to show that $A_q(n,3,2) \geq \lfloor (q-1)n/2 \rfloor$.

If $n$ is even, we apply the strategy outlined in Section III to the codes $\mathcal{C}_1, \ldots, \mathcal{C}_{q-1}$ (i.e., with $s = q-1$) obtained from the one-factorization of $K_n$ above. It is easy to see then that $\mathcal{C}' = \cup_{i=1}^{q-1} \mathcal{C}'_i$ is an $(n,3,2)_q$-code of size $(q-1)n/2$, so $A_q(n,3,2) \geq (q-1)n/2 = \lfloor (q-1)n/2 \rfloor$.

If $n$ is odd, for each $1 \leq i \leq q-1 \leq n-1$, let $\mathcal{C}'_i$ be obtained from $\mathcal{C}_i$ as outlined in Section III, where $\mathcal{C}_i$ ($1 \leq i \leq q-1$) come from the near-one-factorization of $K_n$ above. Further, let

$$\mathcal{C}'_q = \begin{cases} \{(1,2,0,\ldots,0),(0,0,3,4,0,\ldots,0), & \text{if } q \text{ is odd} \\ \ldots,(0,\ldots,0,q-2,q-1,0,\ldots,0)\} & \\ \{(1,2,0,\ldots,0),(0,0,3,4,0,\ldots,0), & \text{if } q \text{ is even.} \\ \ldots,(0,\ldots,0,q-3,q-2,0,\ldots,0)\} & \end{cases}$$

It is easy to verify that $\cup_{i=1}^{q} \mathcal{C}'_i$ is an $(n,3,2)_q$-code of size

$$(q-1)(n-1)/2 + \lfloor (q-1)/2 \rfloor = \lfloor (q-1)n/2 \rfloor$$

that is, $A_q(n,3,2) \geq \lfloor (q-1)n/2 \rfloor$.

Next, we assume that $q > n$.

Obviously, two distinct codewords in an $(n,3,2)_q$-code cannot have the same support, as the distance between them would otherwise be at most two. Hence, it follows that $A_q(n,3,2) \leq \binom{n}{2}$. It therefore suffices to show that $A_q(n,3,2) \geq \binom{n}{2}$.

When $n$ is even, applying the strategy of Section III, with $s = n-1$, shows that $\mathcal{C}' = \cup_{i=1}^{n-1} \mathcal{C}'_i$ is an $(n,3,2)_q$-code of size $(n-1)n/2 = \binom{n}{2}$.

If $n$ is odd, it follows from an application of the strategy of Section III, with $s = n$, that $\cup_{i=1}^{n} \mathcal{C}'_i$ is an $(n,3,2)_q$-code of size $n(n-1)/2 = \binom{n}{2}$.

This completes the proof of Theorem 9. □

## V. EXISTENCE OF DISJOINT $\{3\}$-GDDS OF TYPE $5^1 1^{6t}$

In this section, we prove the existence of a pair of disjoint $\{3\}$-GDDs of type $5^1 1^{6t}$, which is needed in constructing optimal $(n,4,3)_3$-codes in the next section.

*Lemma 6:* There exists a pair of disjoint $\{3\}$-GDDs of type $5^1 1^{6t}$ for $t \in \{0,1,2\}$.

*Proof:* The case $t=0$ is trivial, since a $\{3\}$-GDD of type $5^1$ can have no blocks. A pair of disjoint $\{3\}$-GDDs of type $5^1 1^{6t}$ (on point set $\{0, \ldots, 6t+4\}$ and having $\{0,1,2,3,4\}$ as the group of size five) is given in Tables I and II, for each $t \in \{1,2\}$. □

### A. The Case $t \equiv 0$ or $2 \pmod 3$

Assume $t \geq 3$, since $t \in \{0,2\}$ has been dealt with in Lemma 6.

Let $X = (\mathbb{Z}_{2t+1} \times \mathbb{Z}_3) \cup \{\infty_1, \infty_2\}$ and let $\mathcal{G} = \{G_0, G_1, G_2\}$, where $G_i = \{0, \ldots, 2t\} \times \{i\}$, $i \in \{0,1,2\}$. Let $(X, \mathcal{G}, \mathcal{A})$ and $(X, \mathcal{G}, \mathcal{B})$ be a pair of $\{3\}$-GDDs of type $(2t+1)^3$ with intersection size one, which exists by Theorem 3. Without loss of generality, assume

$$T = \{(0,0),(0,1),(0,2)\} \in \mathcal{A} \cap \mathcal{B}.$$

Let $X_i = G_i \cup \{\infty_1, \infty_2\}$, $i \in \{0,1,2\}$; whence, $|X_i| \equiv 1$ or $3 \pmod 6$. Let $(X_i, \mathcal{A}_i)$ and $(X_i, \mathcal{B}_i)$ be a pair of disjoint $\{3\}$-GDDs of type $3^1 1^{2t}$ with $\{(0,i), \infty_1, \infty_2\}$ as the group of size three, $i \in \{0,1,2\}$. Such a pair of GDDs exists by Theorem 4.

Define

$$G' = \{(0,0),(0,1),(0,2),\infty_1,\infty_2\}$$
$$\mathcal{A}' = (\mathcal{A} \setminus \{T\}) \cup \mathcal{A}_0 \cup \mathcal{A}_1 \cup \mathcal{A}_2$$
$$\mathcal{B}' = (\mathcal{B} \setminus \{T\}) \cup \mathcal{B}_0 \cup \mathcal{B}_1 \cup \mathcal{B}_2.$$

Then $(X, \mathcal{G}', \mathcal{A}')$ and $(X, \mathcal{G}', \mathcal{B}')$ are $\{3\}$-GDDs of type $5^1 1^{6t}$ with $G'$ as the group of size five. It is easy to see that $\mathcal{A}' \cap \mathcal{B}' = \emptyset$. This establishes the following.

*Lemma 7:* There exists a pair of disjoint $\{3\}$-GDDs of type $5^1 1^{6t}$ for all $t \equiv 0$ or $2 \pmod 3$, $t \geq 0$.

Solving the case $t \equiv 1 \pmod{3}$ requires some disjoint incomplete GDDs whose existence we first prove in the next subsection.

### B. Disjoint Incomplete GDDs

Let $n \geq 6$ and let $L$ be a Latin square of side $n$ containing a subsquare $S$ of side three, which exists by Lemma 3. Without loss of generality, assume that $S$ is the $3 \times 3$ subsquare in the top left-hand corner of $L$, and that the entries of $S$ are from the set $[3]$.

Let $\sigma_1 : [3] \to [3]$ and $\sigma_2 : [n] \setminus [3] \to [n] \setminus [3]$ be fixed-point free permutations. Define $\sigma : [n] \to [n]$ such that

$$\sigma(i) = \begin{cases} \sigma_1(i), & \text{if } i \in [3] \\ \sigma_2(i), & \text{if } i \in [n] \setminus [3]. \end{cases}$$

Let $L'$ be the array obtained from $L$ as follows:
1) replace every entry $i$ of $L$ by $\sigma(i)$ and
2) replace the $3 \times 3$ subsquare in the top left-hand corner by $S$.

The $L'$ so obtained is a Latin square of side $n$ containing a subsquare of side three, namely, $S$, in the top left-hand corner. It is obvious from our construction that for all $i, j \in [n]$, the $(i,j)$th entries of $L \setminus S$ and $L' \setminus S$ are different. This proves the following.

*Lemma 8:* For every $n \geq 6$, there exist two disjoint Latin squares of side $n$ with a common subsquare of side three.

It is well known that a Latin square of side $n$ with a fixed subsquare of order $k$ is equivalent to a $\{3\}$-IGDD of type $(n,k)^3$ (see, for example, [44]). The standard construction showing this equivalence when applied with Lemma 8 gives the following result.

*Lemma 9:* For every $n \geq 6$, there exists a pair of disjoint $\{3\}$-IGDDs of type $(n,3)^3$.

### C. The Case $t \equiv 1 \pmod{3}$

Our proof is by induction on $t$. Assume $t \geq 3$ (the case $t \in \{0,1,2\}$ is handled by Lemma 6). Let $X = (\mathbb{Z}_{2t+1} \times \mathbb{Z}_3) \cup \{\infty_1, \infty_2\}$ and let $\mathcal{G} = \{G_0, G_1, G_2\}$, where $G_i = \{0, \ldots, 2t\} \times \{i\}$, $i \in \{0,1,2\}$. Let $X_i = G_i \cup \{\infty_1, \infty_2\}$, $i \in \{0,1,2\}$; whence $|X_i| \equiv 5 \pmod{6}$. By the inductive hypothesis, there exist disjoint $\{3\}$-GDDs, $(X_i, \mathcal{G}_i, \mathcal{A}_i)$, and $(X_i, \mathcal{G}_i, \mathcal{B}_i)$, of type $5^1 1^{2t-2}$, $i \in \{0,1,2\}$. Without loss of generality, assume the group of size five, $F_i \in \mathcal{G}_i$, is given by

$$F_i = \{(0,i),(1,i),(2,i),\infty_1,\infty_2\}, \quad i \in \{0,1,2\}.$$

Further, let $(X, \mathcal{G}, \mathcal{A})$ and $(X, \mathcal{G}, \mathcal{B})$ be two disjoint $\{3\}$-IGDDs of type $(2t+1, 3)^3$, with holes

$$\{(0,0),(1,0),(2,0)\}, \{(0,1),(1,1),(2,1)\},$$
$$\text{and} \{(0,2),(1,2),(2,2)\}.$$

Such a pair of IGDDs exists by Lemma 9.

Define

$$G' = \bigcup_{i \in \{0,1,2\}} F_i$$
$$\mathcal{A}' = \mathcal{A} \cup \mathcal{A}_0 \cup \mathcal{A}_1 \cup \mathcal{A}_2$$
$$\mathcal{B}' = \mathcal{B} \cup \mathcal{B}_0 \cup \mathcal{B}_1 \cup \mathcal{B}_2.$$

Then $(X, \mathcal{G}', \mathcal{A}')$ and $(X, \mathcal{G}', \mathcal{B}')$ are disjoint $\{3\}$-GDDs of type $11^1 16^{6t-6}$, where $G'$ is the group of size 11. Let $(G', \mathcal{G}'', \mathcal{A}'')$, and $(G', \mathcal{G}'', \mathcal{B}'')$ be a pair of disjoint $\{3\}$-GDDs of type $5^1 1^6$, whose existence is provided by Lemma 6. Now, $(X, (\mathcal{G}' \setminus \{G'\}) \cup \mathcal{G}'', \mathcal{A}' \cup \mathcal{A}'')$ and $(X, (\mathcal{G}' \setminus \{G'\}) \cup \mathcal{G}'', \mathcal{B}' \cup \mathcal{B}'')$ are then two disjoint $\{3\}$-GDDs of type $5^1 1^{6t}$.

The above results, together with those in the preceding subsections, give the following.

*Lemma 10:* For every $t \geq 0$, there exists a pair of disjoint $\{3\}$-GDDs of type $5^1 1^{6t}$.

## VI. $A_q(n,4,3)$

In this section, we apply the general strategy of Section III to study $A_q(n,4,3)$. We determine the exact values of $A_q(n,4,3)$ for all $n \equiv 0, 1, 2,$ or $3 \pmod 6$, as well as $A_3(n,4,3)$ for all $n$. For the remaining cases, our method yields lower bounds for $A_q(n,4,3)$ which are fairly close to a known upper bound, as well as some other exact values of $A_q(q,4,3)$, $A_q(q+1,4,3)$, $A_q(q+1,4,4)$, and $A_q(q+2,4,4)$. Asymptotically optimal $(n,4,3)_q$-codes are also constructed.

### A. Upper Bounds for $A_q(n,4,3)$

The following upper bound for $(n, 2w, w)_q$-codes is known.

*Lemma 11 (Fu et al. [3]):* $A_q(n, 2w, w) = \lfloor n/w \rfloor$.

From Lemmas 2 and 11, we infer the following.

*Corollary 1:*

$$A_q(n,4,3) \leq \left\lfloor \frac{(q-1)n}{3} \left\lfloor \frac{n-1}{2} \right\rfloor \right\rfloor =: U(n,q).$$

Let $B(n,q) = (q-1)n(n-1)/6$. Table III gives $U(n,q)$ for values of $n$ modulo six and $q$ modulo three.

When $n \equiv 5 \pmod 6$ and $q \not\equiv 1 \pmod 3$, we have the following better bound.

*Lemma 12:* Let $n \equiv 5 \pmod 6$ and $q \not\equiv 1 \pmod 3$. Then $A_q(n,4,3) \leq U(n,q) - 1$.

*Proof:* Let $n = 6t+5$. Assume first that $q \equiv 0 \pmod 3$. Then Lemma 1 gives

$$A_q(6t+5,4,3)$$
$$\leq \left\lfloor \frac{6t+5}{6t+2} A_q(6t+4,4,3) \right\rfloor$$
$$\leq \left\lfloor \frac{6t+5}{6t+2} U(6t+4,q) \right\rfloor$$
$$= \left\lfloor \frac{6t+5}{6t+2} \left( B(6t+4,q) - \frac{(q-1)(6t+4)}{6} - \frac{2}{3} \right) \right\rfloor$$

$$= \left\lfloor \frac{6t+5}{6t+2}\left(\frac{(q-1)(6t+4)(6t+2)-4}{6}\right)\right\rfloor$$
$$= \left\lfloor (q-1)(6t^2+9t+3) + \frac{q}{3} - 1 - \frac{1}{3t+1}\right\rfloor$$
$$= (q-1)(6t^2+9t+3) + \frac{q}{3} - 2$$
$$= B(n,q) - \frac{5}{3}$$
$$= U(n,q) - 1.$$

The proof in the case $q \equiv 2 \pmod 3$ is similar. However, when $q \equiv 1 \pmod 3$, we only obtain $U(n,q)$ as the upper bound. $\square$

### B. Small Values

In this subsection, we determine $A_q(n,4,3)$ for some small values of $n$ and $q$ that are not covered by the general method described below.

*Lemma 13:* $A_q(7,4,3) = 7(q-1)$ for $q \in \{4,5,6\}$.

*Proof:* Corollary 1 gives $A_q(7,4,3) \leq 7(q-1)$. Let
$$\mathsf{u}_1 = 0000121$$
$$\mathsf{u}_2 = 0033001$$
$$\mathsf{u}_3 = 0020302$$
$$\mathsf{u}_4 = 0004404$$
$$\mathsf{u}_5 = 0005055$$

and let $C_4 = \{\mathsf{u}_1, \mathsf{u}_2, \mathsf{u}_3\}$, $C_5 = C_4 \cup \{\mathsf{u}_4\}$, $C_6 = C_5 \cup \{\mathsf{u}_5\}$. Then, for $q \in \{4,5,6\}$, the set of all cyclic shifts of the codewords in $C_q$ gives a $(7,4,3)_q$-code with $7(q-1)$ codewords, showing $A_q(7,4,3) = 7(q-1)$. $\square$

*Lemma 14:* $A_q(6,4,3) = 4(q-1)$ for $q \in \{4,5,6\}$.

*Proof:* Corollary 1 gives $A_q(6,4,3) \leq 4(q-1)$. The $(6,4,3)_q$-codes showing $A_q(6,4,3) \geq 4(q-1)$ for $q \in \{4,5,6\}$ are the 0-shortened $(7,4,3)_q$-codes in the proof of Lemma 13 above. $\square$

### C. Lower Bounds for $A_q(n,4,3)$ From the General Strategy

We obtain lower bounds for $A_q(n,4,3)$ through explicit constructions. The main strategy is the one outlined in Section III, where the $C_i$ are optimal $(n,4,3)_2$-codes.

We first recall the constructions for optimal $(n,4,3)_2$-codes.
1) For $n \equiv 1 \text{ or } 3 \pmod 6$, the code of an $\text{STS}(n)$ is an optimal $(n,4,3)_2$-code. In this case, $A_2(n,4,3) = n(n-1)/6$.
2) For $n \equiv 0$ or $2 \pmod 6$, let $(X, \mathcal{A})$ be an $\text{STS}(n+1)$ and consider any point $x \in X$. Then the code of $(Y, \mathcal{B})$, where $Y = X \setminus \{x\}$ and $\mathcal{B} = \mathcal{A} \setminus \{A \in \mathcal{A} : x \in A\}$, is an optimal $(n,4,3)_2$-code. In this case, $A_2(n,4,3) = n(n-2)/6$.
3) For $n \equiv 5 \pmod 6$, let $(X, \mathcal{G}, \mathcal{A})$ be a $\{3\}$-GDD of type $5^1 1^{n-5}$ such that $G = \{1,2,3,4,5\}$ is the group of size five. Then the code of $(X, \mathcal{B})$, where $\mathcal{B} = \mathcal{A} \cup \{\{1,2,3\}, \{1,4,5\}\}$, is an optimal $(n,4,3)_2$-code. In this case, $A_2(n,4,3) = (n(n-1)-8)/6$.
4) For $n \equiv 4 \pmod 6$, let $(X, \mathcal{B})$ be the set system above of order $n+1$ whose code is an optimal $(n+1,4,3)_2$-code. Then $(Y, \mathcal{C})$, where $Y = X \setminus \{2\}$ and $\mathcal{C} = \mathcal{B} \setminus \{B \in \mathcal{B} : 2 \in B\}$, is an optimal $(n,4,3)_2$-code. In this case, $A_2(n,4,3) = (n(n-2)-2)/6$.

In order to apply the strategy of Section III, we need to be assured of the existence of $s$ pairwise disjoint optimal $(n,4,3)_2$-codes that satisfy the above condition, for $s$ as large as possible. In particular, if $(X, \mathcal{A}_1), (X, \mathcal{A}_2), \ldots, (X, \mathcal{A}_s)$ are pairwise disjoint set systems giving optimal $(n,4,3)_2$-codes, then their codes $\mathcal{C}_1, \mathcal{C}_2, \ldots, \mathcal{C}_s$ have the above property.

When $n \equiv 1$ or $3 \pmod 6$, $(n,q) \notin \{(7,4),(7,5),(7,6)\}$, by Theorem 5, there exist $q-1$ pairwise disjoint $\text{STS}(n)$ for all $q \leq n-1$. Applying the strategy of Section III, with $s = q-1$, yields

$$A_q(n,4,3) \geq (q-1)A_2(n,4,3) = B(n,q).$$

When $(n,q) \in \{(7,4),(7,5),(7,6)\}$, $A_q(n,4,3) = B(n,q)$ by Lemma 13.

For $n \equiv 0$ or $2 \pmod 6$, $(n,q) \notin \{(6,4),(6,5),(6,6)\}$, Theorem 5 guarantees the existence of $q-1$ pairwise disjoint $\text{STS}(n+1)$ for all $q \leq n$. Starting with these $q-1$ pairwise disjoint $\text{STS}(n+1)$, remove a point, together with all blocks containing that point (there are $n/2$ such blocks in each $\text{STS}(n+1)$). This gives pairwise disjoint binary codes $\mathcal{C}_1, \mathcal{C}_2, \ldots, \mathcal{C}_{n-1}$, each of size $n(n-2)/6$. Applying the strategy of Section III, with $s = q-1$, we have that, for $q \leq n$, $(n,q) \notin \{(6,4),(6,5),(6,6)\}$

$$A_q(n,4,3) \geq (q-1)A_2(n,4,3) = B(n,q) - \frac{(q-1)n}{6}.$$

For $(n,q) \in \{(6,4),(6,5),(6,6)\}$, $A_q(n,4,3) = B(n,q) - (q-1)n/6$ follows from Lemma 14.

For $n \equiv 5 \pmod 6$, we take a large set of maximum $(2,3,n)$-packings, the existence of which is given in Theorem 6. The strategy of Section III, with $s = q-1$, shows that, for $q \leq n-3$

$$A_q(n,4,3) \geq B(n,q) - \frac{4(q-1)}{3}.$$

For $n \equiv 4 \pmod 6$, Theorem 6 gives a large set of maximum $(2,3,n+1)$-packings, and hence $n-3$ disjoint optimal $(n+1,4,3)_2$-codes. Apply the strategy of Section III with $s = q-1$ to get an $(n+1,4,3)_q$-code. Shorten this code by puncturing at a coordinate and remove those codewords whose supports contain this coordinate, in a way similar to the construction of the optimal $(n,4,3)_2$-code above. We choose the coordinate contained in $(n-2)/2$ supports of the first optimal $(n+1,4,3)_2$-code so that there are at most $(n-2)/2 + (q-2)n/2 = (q-1)n/2 - 1$ codewords removed. Then, for $q \leq n-2$, we have

$$A_q(n,4,3) \geq (q-1)A_2(n+1,4,3) - \frac{(q-1)n}{2} + 1$$
$$= B(n+1,q) - \frac{4(q-1)}{3} - \frac{(q-1)n}{2} + 1$$
$$= B(n,q) - \frac{4(q-1)}{3} - \frac{(q-1)n}{6} + 1.$$

By combining the above lower bounds and the upper bounds in Corollary 1 and Lemma 12, and using the values of $U(n,q)$ in Table III, we get the following theorem.

TABLE III
VALUES OF $U(n,q)$ FOR VALUES OF $n$ MODULO SIX AND $q$ MODULO THREE

| $q \pmod 3$ <br> $n \pmod 6$ | 0 | 1 | 2 |
|---|---|---|---|
| 0 | $B(n,q) - \frac{(q-1)n}{6}$ | $B(n,q) - \frac{(q-1)n}{6}$ | $B(n,q) - \frac{(q-1)n}{6}$ |
| 1 | $B(n,q)$ | $B(n,q)$ | $B(n,q)$ |
| 2 | $B(n,q) - \frac{(q-1)n}{6}$ | $B(n,q) - \frac{(q-1)n}{6}$ | $B(n,q) - \frac{(q-1)n}{6}$ |
| 3 | $B(n,q)$ | $B(n,q)$ | $B(n,q)$ |
| 4 | $B(n,q) - \frac{(q-1)n}{6} - \frac{2}{3}$ | $B(n,q) - \frac{(q-1)n}{6}$ | $B(n,q) - \frac{(q-1)n}{6} - \frac{1}{3}$ |
| 5 | $B(n,q) - \frac{2}{3}$ | $B(n,q)$ | $B(n,q) - \frac{1}{3}$ |

*Theorem 10:*

i) For $n \equiv 1$ or $3 \pmod 6$ and $q \leq n - 1$, we have

$$A_q(n,4,3) = \frac{(q-1)n(n-1)}{6}.$$

ii) For $n \equiv 0$ or $2 \pmod 6$ and $q \leq n$, we have

$$A_q(n,4,3) = \frac{(q-1)n(n-2)}{6}.$$

iii) For $n \equiv 5 \pmod 6$ and $q \leq n - 3$, we have

$$U(n,q) - \delta \leq A_q(n,4,3) \leq U(n,q) - \epsilon$$

where

$$\delta = \begin{cases} \frac{4(q-1)-2}{3}, & \text{if } q \equiv 0 \pmod 3 \\ \frac{4(q-1)}{3}, & \text{if } q \equiv 1 \pmod 3 \\ \frac{4(q-1)-1}{3}, & \text{if } q \equiv 2 \pmod 3 \end{cases}$$

and

$$\epsilon = \begin{cases} 1, & \text{if } q \equiv 0 \text{ or } 2 \pmod 3 \\ 0, & \text{if } q \equiv 1 \pmod 3. \end{cases}$$

iv) For $n \equiv 4 \pmod 6$ and $q \leq n - 2$, we have

$$U(n,q) - \delta \leq A_q(n,4,3) \leq U(n,q)$$

where

$$\delta = \begin{cases} \frac{4(q-1)-5}{3}, & \text{if } q \equiv 0 \pmod 3 \\ \frac{4(q-1)-3}{3}, & \text{if } q \equiv 1 \pmod 3 \\ \frac{4(q-1)-4}{3}, & \text{if } q \equiv 2 \pmod 3. \end{cases}$$

Therefore, the construction above yields optimal $(n,4,3)_q$-codes when $n \equiv 0, 1, 2,$ or $3 \pmod 6$ and $q$ is (roughly) $\leq n$.

When $q$ is fixed, Theorem 10 also shows that the construction above yields families of $(n,4,3)_q$-codes that are asymptotically optimal.

*Theorem 11:* For fixed $q$, $A_q(n,4,3) \sim (q-1)n^2/6$.

### D. $A_3(n,4,3)$ for $n \equiv 4$ or $5 \pmod 6$

The results in the previous subsection determine $A_3(n,4,3)$ for all $n \not\equiv 4$ or $5 \pmod 6$. Here, we determine the remaining values of $A_3(n,4,3)$.

Let $n \equiv 5 \pmod 6$ and consider a pair of disjoint $\{3\}$-GDDs $(X, \mathcal{G}, \mathcal{A})$, and $(X, \mathcal{G}, \mathcal{B})$ of type $5^1 1^{n-5}$ with $\{0,1,2,3,4\}$ as the group of size five, which exists by Lemma 10. Let $\mathcal{C_A}$ and $\mathcal{C_B}$ be the codes of $(X, \mathcal{A})$ and $(X, \mathcal{B})$, respectively. $\mathcal{C_A}$ and $\mathcal{C_B}$ are obviously disjoint and we can apply the strategy of Section III to obtain an $(n,4,3)_3$-code $\mathcal{C}'$. Taking all the codewords in $\mathcal{C}'$ together with the codewords of an optimal $(5,4,3)_3$-code (which has five codewords [24]) on the first five coordinates still gives an $(n,4,3)_3$-code since every block in $\mathcal{A}$ and $\mathcal{B}$ intersects the group of size five in only one point.

The size of this $(n,4,3)_3$-code is

$$|\mathcal{A}| + |\mathcal{B}| + 5 = 2\left(\frac{n(n-1)/2 - 10}{3}\right) + 5$$
$$= B(n,3) - \frac{5}{3}$$
$$= U(n,3) - 1$$

which is optimal by Lemma 12.

Now let $n \equiv 4 \pmod 6$ and consider the $(n+1, 4, 3)_3$-code $\mathcal{C}'$ constructed above. Shorten $\mathcal{C}'$ by puncturing at coordinate zero and removing all the codewords whose support contains zero to give an $(n,4,3)_3$-code $\mathcal{C}''$ of size

$$|\mathcal{C}'| - (n-4) = \frac{(n+2)(n-4)}{3}.$$

Taking $\mathcal{C}''$ together with the codewords $01110\cdots 0$ and $20220\cdots 0$ (which form an optimal $(4,4,3)_3$-code on the first four coordinates) gives an $(n,4,3)_3$-code of size

$$\frac{(n+2)(n-4)}{3} + 2 = B(n,3) - \frac{n}{3} - \frac{2}{3}$$
$$= U(n,3)$$

which is therefore optimal.

### E. Improved Upper Bound

While the results in Section VI-C have the constraint that $q$ is (roughly) $\leq n$, the complementary case of relatively large $q$ (with respect to $n$) is in fact much easier.

In fact, for the distance between two $q$-ary words of weight three to be greater than three, their supports cannot be identical. Hence, for any $(n, 4, 3)_q$-code, the supports of the codewords must all be distinct. Clearly, there are at most $\binom{n}{3}$ distinct supports of size three among codewords of length $n$. Hence, Corollary 1 can be refined to the following.

*Theorem 12:* $A_q(n, 4, 3) \leq \min\left\{U(n, q), \binom{n}{3}\right\}$.

It is easy to observe the following.
  i) For $n \equiv 1, 3$, or $5 \pmod{6}$, $q \geq n - 1$ implies that $U(n, q) \geq \binom{n}{3}$, hence $A_q(n, 4, 3) \leq \binom{n}{3}$.
  ii) For $n \equiv 0, 2$, or $4 \pmod{6}$, $q \geq n$ implies that $U(n, q) \geq \binom{n}{3}$, hence $A_q(n, 4, 3) \leq \binom{n}{3}$.

Consider the map $T : \mathcal{H}_2(n, 3) \to \mathbb{Z}_n$ defined by

$$(c_0, c_1, \ldots, c_{n-1}) \mapsto \sum_{i=0}^{n-1} i c_i \pmod{n}.$$

As in the Graham–Sloane construction [45], $\mathcal{C}_i = T^{-1}(i)$ is an $(n, 4, 3)_2$-code for every $1 \leq i \leq n$, and $\cup_{i=1}^{n} \mathcal{C}_i = \mathcal{H}_2(n, 3)$. In particular, $\sum_{i=1}^{n} |\mathcal{C}_i| = \binom{n}{3}$. If $q \geq n + 1$, using the strategy of Section III with $s = n$, it follows that $\cup_{i=1}^{n} \mathcal{C}_i'$ is an optimal $(n, 4, 3)_q$-code, and $A_q(n, 4, 3) = \binom{n}{3}$.

The following theorem summarizes what we know about $A_q(n, 4, 3)$.

*Theorem 13:*
  i) If $n \leq q - 1$, then
  $$A_q(n, 4, 3) = \binom{n}{3}.$$
  ii) For the other values of $n$:
    a) For $n \equiv 1$ or $3 \pmod{6}$, we have
    $$A_q(n, 4, 3) = \begin{cases} \frac{(q-1)n(n-1)}{6}, & \text{if } n \geq q+1; \\ \binom{n}{3}, & \text{if } n = q. \end{cases}$$
    b) For $n \equiv 0$ or $2 \pmod{6}$, we have
    $$A_q(n, 4, 3) = \frac{(q-1)n(n-2)}{6}$$
    if $n \geq q$.
    c) For $n \equiv 4 \pmod{6}$, we have: if $n \geq q + 2$
    $$U(n, q) - \delta \leq A_q(n, 4, 3) \leq U(n, q),$$
    where
    $$\delta = \begin{cases} \frac{4(q-1)-5}{3}, & \text{if } q \equiv 0 \pmod{3} \\ \frac{4(q-1)-3}{3}, & \text{if } q \equiv 1 \pmod{3} \\ \frac{4(q-1)-4}{3}, & \text{if } q \equiv 2 \pmod{3}. \end{cases}$$
    We also have, for $q \in \{n-1, n\}$
    $$A_{n-2}(n, 4, 3) \leq A_q(n, 4, 3) \leq U(n, q).$$
    Moreover, $A_3(n, 4, 3) = U(n, 3)$ for $n \equiv 4 \pmod{6}$.

  d) For $n \equiv 5 \pmod{6}$, we have the following: if $n \geq q + 3$
    $$U(n, q) - \delta \leq A_q(n, 4, 3) \leq U(n, q) - \epsilon,$$
    where
    $$\delta = \begin{cases} \frac{4(q-1)-2}{3}, & \text{if } q \equiv 0 \pmod{3} \\ \frac{4(q-1)}{3}, & \text{if } q \equiv 1 \pmod{3} \\ \frac{4(q-1)-1}{3}, & \text{if } q \equiv 2 \pmod{3} \end{cases}$$
    and
    $$\epsilon = \begin{cases} 1, & \text{if } q \equiv 0 \text{ or } 2 \pmod{3} \\ 0, & \text{if } q \equiv 1 \pmod{3}. \end{cases}$$
    We also have, for $q \in \{n-2, n-1, n\}$
    $$A_{n-3}(n, 4, 3) \leq A_q(n, 4, 3) \leq U(n, q) - \epsilon.$$
    Moreover, $A_3(n, 4, 3) = U(n, 3) - 1$ for $n \equiv 5 \pmod{6}$.

*Proof:* Most of the results were obtained earlier. The remaining cases are as follows.

For $n \equiv 1$ or $3 \pmod{6}$ and $q = n$, the construction earlier using STS$(q)$ already yields an optimal code, so $A_q(q, 4, 3) = \binom{q}{3}$.

For $n \equiv 4 \pmod{6}$, $q \in \{n-1, n\}$, and for $n \equiv 5 \pmod{6}$, $q \in \{n-2, n-1, n\}$, the inequalities are obvious. $\square$

*F. Some Values of $A_q(q, 4, 3)$, $A_q(q+1, 4, 3)$, $A_q(q+1, 4, 4)$, and $A_q(q+2, 4, 4)$*

While the preceding methods fail to determine some values of $A_q(q, 4, 3)$ and $A_q(q+1, 4, 3)$, it is nonetheless possible to obtain these values in some further cases. As a by-product, we also obtain some values of $A_q(q+1, 4, 4)$ and $A_q(q+2, 4, 4)$.

Recall that bounds (NB05) and (NB06) of Ding *et al.* [4], obtained using finite geometries, give, in particular, the following lower bounds.

  i) If $q$ is the power of an odd prime, then
  $$A_q(q+1, 4, 4) \geq \frac{(q+1)q(q-1)^2(q-2)}{4!}. \tag{1}$$
  ii) If $q = 2^h$, then
  $$A_q(q+2, 4, 4) \geq \frac{(q+2)(q+1)q(q-1)^2}{4!}. \tag{2}$$

On the other hand, Lemma 2 shows that

$$A_q(q+1, 4, 4) \leq \frac{(q+1)(q-1)}{4} A_q(q, 4, 3) \tag{3}$$
$$\leq \frac{(q+1)(q-1)q(q-1)(q-2)}{4 \cdot 3!} \tag{4}$$

and

$$A_q(q+2, 4, 4) \leq \frac{(q+2)(q-1)}{4} A_q(q+1, 4, 3) \tag{5}$$
$$\leq \frac{(q+2)(q-1)(q+1)q(q-1)}{4 \cdot 3!}. \tag{6}$$

Theorem 14 then follows from (1)–(6).

*Theorem 14:*

i) If $q$ is the power of an odd prime, then

$$A_q(q+1,4,4) = (q-1)\binom{q+1}{4} \quad \text{and}$$

$$A_q(q,4,3) = \binom{q}{3}.$$

ii) If $q = 2^h$, then

$$A_q(q+2,4,4) = (q-1)\binom{q+2}{4} \quad \text{and}$$

$$A_q(q+1,4,3) = \binom{q+1}{3}.$$

### G. An Alternate Construction of Asymptotically Optimal $(n,4,3)_q$-Codes

In this subsection, we give another way of constructing a family of asymptotically optimal $(n,4,3)_q$-codes.

Graham and Sloane [45] constructed disjoint $(n,4,3)_2$-codes $\mathcal{C}_0, \mathcal{C}_1, \ldots, \mathcal{C}_{n-1}$ such that $\cup_{i=0}^{n-1} \mathcal{C}_i = \mathcal{H}_2(n,3)$. Let $n = (q-1)\tau + r$, where $0 \leq r < q-1$. Hence, $\tau = \lfloor n/(q-1) \rfloor$. Partition $\{\mathcal{C}_0, \mathcal{C}_1, \ldots, \mathcal{C}_{n-1}\}$ arbitrarily into $\tau$ parts, each of size $q-1$, and one part of size $r$.

Discard the part of size $r$. Since an $(n,4,3)_2$-code has size at most $n(n-1)/6$, at most $rn(n-1)/6 \leq (q-2)n(n-1)/6$ codewords have been thrown away. Therefore, the remaining $\tau$ parts of size $q-1$ have at least

$$\binom{n}{3} - (q-2)n(n-1)/6 = n(n-1)(n-q)/6$$

codewords altogether. Hence, there must be one part of size $q-1$ having at least

$$\frac{n(n-1)(n-q)}{6t} = \frac{n(n-1)(n-q)}{6\lfloor n/(q-1) \rfloor}$$
$$\geq \frac{(q-1)(n-q)}{n} \cdot \frac{n(n-1)}{6}$$
$$= (q-1)\left(1 - \frac{q}{n}\right) \cdot \frac{n(n-1)}{6}$$
$$= (1 - o(1))\frac{(q-1)n(n-1)}{6}$$

codewords. Take this part, say $\{\mathcal{C}_1, \mathcal{C}_2, \ldots, \mathcal{C}_{q-1}\}$, and apply the strategy described in Section III. It follows that $\cup_{i=1}^{q-1} \mathcal{C}'_i$ is an $(n,4,3)_q$-code of size $(1 - o(1))B(n,q)$, for all fixed $q$ (where the $o(1)$ is in $n$).

## VII. $A_q(13,6,4)$

The method of Section III can be applied to yield $q$-ary constant-weight codes so long as large sets of certain set systems exist.

By Theorem 7, a large set of 2-$(13,4,1)$ designs exists. Such a large set consists of 55 pairwise disjoint set systems. Moreover, an optimal binary $(13,6,4)_2$-code exists (given by such a design), and $A_2(13,6,4) = 13$. Hence, for $q \leq 56$, the construction of Section III with $s = q-1$, gives

$$A_q(13,6,4) \geq 13(q-1).$$

On the other hand, Lemmas 2 and 11 also yield

$$A_q(13,6,4) \leq 13(q-1).$$

When $q > 56$, then an upper bound for $A_q(13,6,4)$ is given by $\binom{13}{4} = 13 \cdot 55$. Hence, the construction of Section III, with $s = 55$, together with the above large set, gives rise to $A_q(13,6,4) = \binom{13}{4}$ for $q > 56$. Hence we get the following.

*Theorem 15:*

$$A_q(13,6,4) = \begin{cases} 13(q-1), & \text{if } q \leq 56 \\ \binom{13}{4}, & \text{if } q > 56. \end{cases}$$

## VIII. $A_q(n, w+1, w)$

Since the $(n,4,3)_q$-codes studied in Section VI form a special case of $(n, w+1, w)_q$-codes, it is natural to wonder if similar techniques may shed some light on the values of $A_q(n, w+1, w)$ for other values of $w$. This is the question investigated in this section. While we are unable to determine the exact values of $A_q(n, w+1, w)$ for general $w$, some bounds are obtained. When $q = 3$ or $4$, and for $w$ even and $n$ sufficiently large, the lower bound obtained using our method is stronger than the $q$-ary Gilbert–Varshamov bound.

We begin with an upper bound on $A_q(n, w+1, w)$.

### A. An Upper Bound

The following upper bound on $A_q(n, w+1, w)$ holds for all values of $q, n$ and $w$.

*Lemma 15:*

$$A_q(n, w+1, w) \leq (q-1)^{\lceil w/2 \rceil} \binom{n}{\lceil w/2 \rceil} / \binom{w}{\lceil w/2 \rceil}.$$

In particular, when $w$ is even,

$$A_q(n, w+1, w) \leq (q-1)^{w/2} \binom{n}{w/2} / \binom{w}{w/2}.$$

*Proof:* Let $\mathcal{C}$ be an $(n, w+1, w)_q$-code. Clearly, no two codewords can have identical supports. Hence, by letting $X$ be a set of $n$ points corresponding to the $n$ coordinates, and by setting $\mathcal{A}$ to be the set of blocks corresponding to the supports of the codewords in $\mathcal{C}$, it follows that $(X, \mathcal{A})$ is a $\{w\}$-uniform set system. Suppose $T$ is a $\lceil w/2 \rceil$-subset of $X$ and $T$ is contained in more than $(q-1)^{\lceil w/2 \rceil}$ blocks of $\mathcal{A}$. Then $\mathcal{C}$ cannot have distance $w+1$. Hence, every $\lceil w/2 \rceil$-subset of $X$ is contained in at most $(q-1)^{\lceil w/2 \rceil}$ blocks. $(X, \mathcal{A})$ is, therefore, a $\lceil w/2 \rceil$-$(n, w, (q-1)^{\lceil w/2 \rceil})$ packing, giving

$$|\mathcal{A}| \leq (q-1)^{\lceil w/2 \rceil} \binom{n}{\lceil w/2 \rceil} / \binom{w}{\lceil w/2 \rceil}$$

by Lemma 4. □

### B. A Lower Bound for $A_q(n, w+1, w)$

To obtain a lower bound for $A_q(n, w+1, w)$, we first recall the following useful fact on $\{k\}$-uniform set systems.

*Theorem 16 (Ganter, Pelikan, and Teirlinck [46]):* If $(X, \mathcal{A})$ and $(X, \mathcal{B})$ are $\{k\}$-uniform set systems such that $|\mathcal{A}| \cdot |\mathcal{B}| < \binom{|X|}{k}$, then there exists a permutation $\pi : X \to X$ such that $\pi(\mathcal{A}) \cap \mathcal{B} = \emptyset$.

Theorem 16 can be used to ensure the existence of a certain number of pairwise disjoint binary constant-weight codes, which then permits the application of the strategy of Section III.

*Corollary 2:* If there exists an $(n, d, w)_2$-code $\mathcal{C}$ such that $|\mathcal{C}|^s < \binom{n}{w}$, then there exist $s$ pairwise disjoint $(n, d, w)_2$-codes, each of size $|\mathcal{C}|$.

*Proof:* Let $\mathcal{A}$ be the set of supports of the $(n, d, w)_2$-code $\mathcal{C}$. Then if $|\mathcal{C}|^2 = |\mathcal{A}|^2 < \binom{n}{w}$, Theorem 16 implies there exists a permutation $\pi$ such that $\mathcal{A}$ and $\pi(\mathcal{A})$ are disjoint. Now suppose we have $s-1$ pairwise disjoint $(n, d, w)_2$-codes, each of size $|\mathcal{C}|$. By Theorem 16, if $|\mathcal{C}|^{s-1} \cdot |\mathcal{C}| = |\mathcal{C}|^s < \binom{n}{w}$, then we can find an $(n, d, w)_2$-code disjoint from each of the $s-1$ $(n, d, w)_2$-codes. The proof is complete by induction. □

*Lemma 16:* For $1 \leq t < w$, there exist $\lfloor w/t \rfloor$ pairwise disjoint optimal $(n, 2(w-t+1), w)_2$-codes for all $n$ sufficiently large.

*Proof:* By the first Johnson bound

$$A_2(n, 2(w-t+1), w) \leq \binom{n}{t}/\binom{w}{t} = n^t / \binom{w}{t} t! + o(n^t).$$

For any optimal $(n, 2(w-t+1), w)_2$-code $\mathcal{C}$, if $|\mathcal{C}|$ satisfies $|\mathcal{C}|^{\lfloor w/t \rfloor} < \binom{n}{w}$ for $n$ sufficiently large, then the lemma follows from Corollary 2. It remains, therefore, to check that $|\mathcal{C}|^{\lfloor w/t \rfloor} \leq |\mathcal{C}|^{w/t} < \binom{n}{w}$ for $n$ sufficiently large.

It suffices to prove that, for $n$ sufficiently large, we have

$$\left(\frac{n^t}{\binom{w}{t} t!}\right)^{w/t} < \frac{n^w}{w!}$$

or, equivalently

$$w! < \left(\binom{w}{t} t!\right)^{w/t}. \tag{7}$$

We claim that the right-hand side of (7) decreases as $t$ increases. Indeed, for $t < w$, we have

$$\left(\binom{w}{t+1}(t+1)!\right)^{w/(t+1)} = \left(\prod_{i=0}^{t}(w-i)\right)^{w/(t+1)}$$

$$< \left(\left(\prod_{i=0}^{t-1}(w-i)\right)^{1+\frac{1}{t}}\right)^{w/(t+1)}$$

$$= \left(\prod_{i=0}^{t-1}(w-i)\right)^{w/t}$$

$$= \left(\binom{w}{t} t!\right)^{w/t}.$$

Hence, (7) is true, so the lemma follows. □

*Lemma 17:* Let $1 \leq t \leq (w+1)/2$ and $q-1 \leq \lfloor w/t \rfloor$. Then

$$A_q(n, w+1, w) \geq (q-1) A_2(n, 2(w-t+1), w)$$

for all $n$ sufficiently large.

*Proof:* By Lemma 16, there exist $q-1$ pairwise disjoint optimal $(n, 2(w-t+1), w)_2$-codes. The construction of Section III, with $s = q-1$, gives the required $(n, w+1, w)_q$-code. □

The Gilbert–Varshamov bound for $q$-ary constant-weight codes is the following:

$$A_q(n, d, w) \geq \frac{\binom{n}{w}(q-1)^w}{S_{d-1}^{n,w}} \tag{8}$$

where (with $m_i^{n,w} = \min(\lfloor i/2 \rfloor, n-w)$ for each $0 \leq i \leq r$)

$$S_r^{n,w} = \sum_{i=0}^{r} \sum_{j=0}^{m_i^{n,w}} \binom{w}{j}\binom{n-w}{j}\binom{w-j}{i-2j}(q-1)^j(q-2)^{i-2j}$$

is the size of a sphere of radius $r$ in the $q$-ary Johnson space [47].

When $w$ is even, for large $n$ and $d = w + 1$, the Gilbert–Varshamov bound is approximately

$$\frac{\binom{n}{w}(q-1)^w}{S_w^{n,w}} \sim n^{w/2}(q-1)^{w/2}\frac{((w/2)!)^3}{(w!)^2}$$

since

$$S_w^{n,w} \sim \frac{\binom{w}{w/2}(q-1)^{w/2}}{(w/2)!} n^{w/2}.$$

Note that $A_2(n, 2(w-t+1), w)$ is the number of blocks in a maximum $t$-$(n, w, 1)$ packing. Rödl's [48] celebrated proof of the Erdös–Hanani conjecture shows that $A_2(n, 2(w-t+1), w) = (1 - o(1))\binom{n}{t}/\binom{w}{t}$.

*Theorem 17 (Rödl [48]):* Let $(X, \mathcal{A})$ be a maximum $t$-$(n, w, 1)$ packing. Then $|\mathcal{A}| = (1 - o(1))\binom{n}{t}/\binom{w}{t}$.

Hence, for fixed $w \equiv 0 \pmod{2}$ and $q$, Lemma 17 and Theorem 17 with $t = w/2$ yield the following asymptotic lower bound for $A_q(n, w+1, w)$:

$$A_q(n, w+1, w) \gtrsim \frac{(q-1)(w/2)!}{w!} n^{w/2}. \qquad (9)$$

It is easy to show, by induction, that, for $q \in \{3, 4\}$

$$\binom{w}{w/2} > (q-1)^{w/2-1}.$$

Hence, for $q = 3$ or $4$

$$\frac{2(w/2)!}{w!} > (q-1)^{w/2} \frac{((w/2)!)^3}{(w!)^2}.$$

This means that Lemma 17 gives a better asymptotic lower bound than the Gilbert–Varshamov bound when $w$ is even, for $q \in \{3, 4\}$. Therefore, for $q \in \{3, 4\}$, the codes constructed in this section beat the Gilbert–Varshamov bound asymptotically when $w$ is even.

Lemma 15 and (9) yield the following.

*Corollary 3:* For $q \in \{3, 4\}$ and $w$ even

$$(q-1)V(n, w) \lesssim A_q(n, w+1, w) \lesssim (q-1)^{w/2} V(n, w)$$

where $V(n, w) = \frac{(w/2)!}{w!} n^{w/2}$. This lower bound is better than the Gilbert–Varshamov bound.

## IX. A Probabilistic Construction for $(n, d, w)_q$-Codes

In this section, we turn our attention to a probabilistic construction of $(n, d, w)_q$-codes. For large values of $q$, the codes obtained via this construction can have sizes close to a known upper bound for $A_q(n, d, w)$. The method applies for all $d \leq 2w$.

Let $t = \lceil (2w - d + 1)/2 \rceil$. Consider a $t$-$(n, w, \lambda)$ packing $(X, \mathcal{B})$. To each block $B \in \mathcal{B}$, we associate a $q$-ary codeword $\mathbf{c}_B$ of length $n$ in the following way: the coordinates of the codeword are indexed by the points in $X$, the support of $\mathbf{c}_B$ corresponds to precisely the points lying on $B$, and every nonzero coordinate of $\mathbf{c}_B$ is assigned a random value from $\{1, 2, \ldots, q-1\}$ with equal probability $1/(q-1)$. Let $\mathcal{C}$ denote the $q$-ary code thus obtained.

For each codeword $\mathbf{u} \in \mathcal{C}$, the *conflicts* of $\mathbf{u}$, denoted $\mathrm{conf}(\mathbf{u})$, are the set of codewords $\mathbf{v} \in \mathcal{C} \setminus \{\mathbf{u}\}$ such that $d_H(\mathbf{u}, \mathbf{v}) < d$. A necessary condition for $\mathbf{v} \in \mathrm{conf}(\mathbf{u})$ is that the supports of $\mathbf{u}$ and $\mathbf{v}$ must intersect at $\geq t = \lceil (2w - d + 1)/2 \rceil$ coordinates. There are at most $\binom{w}{t}(\lambda - 1)$ other codewords in $\mathcal{C}$ whose support intersects the support of $\mathbf{u}$ at $t$ points. For any of these codewords $\mathbf{v}$, in order to have $d_H(\mathbf{u}, \mathbf{v}) < d$, at least $\lfloor (2w-d+1)/2 \rfloor$ of the $t$ coordinates that $\mathbf{u}$ and $\mathbf{v}$ have in common need to be identical. Therefore

$$\mathbb{E}(|\mathrm{conf}(\mathbf{u})|) \leq \binom{w}{t}(\lambda - 1)\frac{\binom{t}{\lfloor (2w-d+1)/2 \rfloor}}{(q-1)^{\lfloor (2w-d+1)/2 \rfloor}}$$

where $\mathbb{E}$ denotes the expectation.

By linearity of expectation, since $|\mathcal{B}| \leq \lambda\binom{n}{t}/\binom{w}{t}$, we have

$$\mathbb{E}\left(\sum_{\mathbf{u} \in \mathcal{C}} |\mathrm{conf}(\mathbf{u})|\right) \leq \binom{w}{t}\frac{(\lambda-1)\binom{t}{\lfloor (2w-d+1)/2 \rfloor}}{(q-1)^{\lfloor (2w-d+1)/2 \rfloor}} \cdot \frac{\lambda\binom{n}{t}}{\binom{w}{t}}$$

$$= \frac{\lambda(\lambda-1)\binom{t}{\lfloor (2w-d+1)/2 \rfloor}}{(q-1)^{\lfloor (2w-d+1)/2 \rfloor}}\binom{n}{t}.$$

Therefore, there exists a construction of $\mathcal{C}$ resulting in at most

$$\frac{\lambda(\lambda-1)\binom{t}{\lfloor (2w-d+1)/2 \rfloor}}{(q-1)^{\lfloor (2w-d+1)/2 \rfloor}}\binom{n}{t}$$

conflicts. Deleting these conflicts gives an $(n, d, w)_q$-code with at least

$$\left(\frac{\lambda}{\binom{w}{t}} - \frac{\lambda(\lambda-1)\binom{t}{\lfloor (2w-d+1)/2 \rfloor}}{(q-1)^{\lfloor (2w-d+1)/2 \rfloor}}\right)\binom{n}{t} \qquad (10)$$

codewords. For any $\epsilon > 0$, taking $\lambda = q^{\lfloor (2w-d+1)/2 \rfloor - \epsilon}$ gives an $(n, d, w)_q$-code with at least

$$(1 - o(1))\frac{q^{\lfloor (2w-d+1)/2 \rfloor - \epsilon}\binom{n}{t}}{\binom{w}{t}}$$

codewords, where the $o(1)$ is with respect to $q$. In other words, for $q$ large

$$A_q(n, d, w) \geq \begin{cases} (1 - o(1))\frac{q^{t-\epsilon}\binom{n}{t}}{\binom{w}{t}}, & \text{if } d \text{ is odd} \\ (1 - o(1))\frac{q^{t-1-\epsilon}\binom{n}{t}}{\binom{w}{t}}, & \text{if } d \text{ is even.} \end{cases}$$

On the other hand, Lemma 2 gives

$$A_q(n, d, w) \leq \begin{cases} \frac{(q-1)^t\binom{n}{t}}{\binom{w}{t}}, & \text{if } d \text{ is odd} \\ \frac{(q-1)^{t-1}\binom{n}{t}}{\binom{w}{t}}, & \text{if } d \text{ is even.} \end{cases}$$

*Theorem 18:* Given positive integers $q, n, d, w$ such that $d \leq 2w$, and $w \leq n$, let $t = \lceil (2w - d + 1)/2 \rceil$. For any real number $\epsilon > 0$ and $q$ sufficiently large, we have

i) if $d$ is odd

$$(1 - o(1))\frac{q^{t-\epsilon}\binom{n}{t}}{\binom{w}{t}} \leq A_q(n, d, w) \leq (1 - o(1))\frac{q^t\binom{n}{t}}{\binom{w}{t}};$$

ii) if $d$ is even

$$(1 - o(1))\frac{q^{t-1-\epsilon}\binom{n}{t}}{\binom{w}{t}} \leq A_q(n, d, w) \leq (1 - o(1))\frac{q^{t-1}\binom{n}{t}}{\binom{w}{t}}.$$

Alternatively, let

$$\lambda \approx \begin{cases} \frac{1}{2}\left(\frac{(q-1)^t}{\binom{w}{t}} + 1\right), & \text{if } d \text{ is odd} \\ \frac{1}{2}\left(\frac{(q-1)^{t-1}}{\binom{w}{t}t} + 1\right), & \text{if } d \text{ is even.} \end{cases}$$

Then, (10) gives the following result.

*Theorem 19:* Given positive integers $q, n, d, w$, such that $d \leq 2w$ and $w \leq n$, we have

i) if $d$ is odd,

$$\frac{1}{4\binom{w}{t}} \cdot \frac{(q-1)^t \binom{n}{t}}{\binom{w}{t}} \lesssim A_q(n,d,w) \leq \frac{(q-1)^t \binom{n}{t}}{\binom{w}{t}};$$

ii) if $d$ is even

$$\frac{1}{4\binom{w}{t}t} \cdot \frac{(q-1)^{t-1} \binom{n}{t}}{\binom{w}{t}} \lesssim A_q(n,d,w) \leq \frac{(q-1)^{t-1} \binom{n}{t}}{\binom{w}{t}}$$

where $t = \lceil (2w-d+1)/2 \rceil$.

## X. CONCLUSION

In this paper, we introduced a general combinatorial construction for $q$-ary constant-weight codes and used it to derive the exact values of $A_q(n,d,w)$ for several infinite families of $(n,d,w,q)$ parameter sets. Improved general bounds were also obtained on the size of optimal $(n,d,w)_q$-codes.

One interesting aspect of this research is that it reveals the intimate connection between $q$-ary constant-weight codes and sets of pairwise disjoint combinatorial designs of various types, thus suggesting new problems and application areas for combinatorial design theory. One such problem is as follows.

*Problem 1:* Determine the existence of $m$ pairwise disjoint $\{3\}$-GDDs (having common groups) of type $g^t 1^r$.

The case $m = 1$ has been solved by Colbourn *et al.* [49]. The case $m = 2$ is solved for $r = 0$ by Butler and Hoffman [28], for $g = 3$ by Chee [29], and for $(g,t) \in \{(5,1), (11,1)\}$ in this paper. Further progress on this problem would be interesting.

### ACKNOWLEDGMENT

The authors thank the anonymous reviewers for their comments and are especially grateful to the reviewer who supplied the elegant cyclic codes that significantly shorten the proofs of Lemmas 13 and 14.